\documentclass[aps,twocolumn,nofootinbib,
preprintnumbers,superscriptaddress]{revtex4-1}
\usepackage{latexsym}
\usepackage{graphics}
\usepackage{amsmath}
\usepackage{amsfonts}
\renewcommand{\vec}[1]{{\mathbf{#1}}}
\usepackage{amsmath}
\usepackage{amssymb}
\usepackage[paper=letterpaper,margin=1in]{geometry}
\usepackage{braket}
\usepackage{upgreek}

\newcommand{\beq}{\begin{eqnarray}}
\newcommand{\eeq}{\end{eqnarray}}

\renewcommand{\vec}[1]{\boldsymbol{#1}}

\def\t{\tau}

\def\det{{\rm{det}}}

\usepackage{tikz}

\newtheorem{thm}{Theorem}[section]

\newtheorem{preremark}[thm]{Remark}

\begin{document}
\title{ Discrete Symmetry Breaking Defines the Mott Quartic Fixed Point}
\author{Edwin W. Huang}
\thanks{All authors contributed equally.}
\affiliation{Department of Physics and Institute for Condensed Matter Theory,
University of Illinois, 1110 W. Green Street, Urbana, IL 61801}
\author{Gabriele La Nave}
\thanks{All authors contributed equally.}
\affiliation{Department of Mathematics, University of Illinois,
Urbana, Il. 61820}
\author{Philip W. Phillips}
\thanks{All authors contributed equally.}
\affiliation{Department of Physics and Institute for Condensed Matter Theory,
University of Illinois, 1110 W. Green Street, Urbana, IL 61801}

\begin{abstract}
{Because Fermi liquids are inherently non-interacting states of matter, all electronic levels below the chemical potential are doubly occupied.  Consequently, the simplest way of breaking Fermi liquid theory is to engineer a model in which some of those states are singly occupied keeping time-reversal invariance intact.    We show that breaking an overlooked\cite{AHO4} local-in-momentum space $\mathbb Z_2$ symmetry of a Fermi liquid does precisely this.  As a result, while the Mott transition from a Fermi liquid is correctly believed to obtain without the breaking of any continuous symmetry, a discrete  symmetry is broken. This symmetry breaking serves as an organizing principle for Mott physics whether it arises from the tractable Hatsugai-Kohmoto (HK) model or the intractable Hubbard model.  That both are controlled by the same fixed point we establish through a renormalization group analysis.  An experimental manifestation of this fixed point is the onset of  particle-hole asymmetry, a widely observed\cite{chen,sawatzky,rmpphillips,kohno,singhpha,jwilkins,pickettpha,phs,ong} phenomenon in strongly correlated systems.  Theoretically, the singly-occupied region of the spectrum gives rise to a surface of zeros of the single-particle Green function, denoted as the Luttinger surface.  Using K-homology, we show that the Bott topological invariant guarantees the stability of this surface to local perturbations.  Our proof demonstrates that the strongly coupled fixed point only corresponds to those Luttinger surfaces with co-dimension $p+1$ with $p$ odd. We conclude that the Hubbard and HK models both lie in the same high temperature universality class and are controlled by the broken $\mathbb Z_2$ symmetry quartic fixed point.}\end{abstract}

\maketitle

Symmetry is a fundamental organizing principle of nature.  A case in point is the simplest example of symmetry, namely permutations.   This symmetry helps organize identical fundamental particles into two groups: fermions,  odd under interchange and bosons, even under permutation.  Since the permutation group has a finite number of elements, $\pm 1$, it is an example of a discrete symmetry.  What we show here is that a group as simple as the permutation group, namely $\mathbb Z_2$, controls the transition from a non-interacting collection of electrons constituting a Fermi surface to a state that strongly violates the traditional theory of metals, namely the Mott paramagnetic state which insulates although the band is half full.  The Fermi surface retains $\mathbb Z_2$ symmetry but the Mott state does not.  

A manifestation of this symmetry breaking  is the resultant asymmetry upon particle-hole addition or removal,  {that is upon doping}.  In a non-interacting electron system, adding or subtracting an electron is a symmetrical process.  However, cuprate superconductors as varied as underdoped Bi$_2$Sr$_2$CaCu$_2$O$_{8+\delta}$ (Bi-2212) and Ca$_{2-x}$Na$_x$CuO$_2$Cl$_2$ (Na-CCOC) all exhibit scanning tunneling spectra\cite{davis2004,lanzarapha,phs,ong}  with a distinct asymmetry in terms of particle addition and removal.  The cuprates are not alone here as there are numerous electronic systems\cite{singhpha,jwilkins,pickettpha} which exhibit  particle-hole asymmetry at low energies upon the addition or removal of an electron. Although it is now commonplace to attribute particle-hole asymmetry to strong correlations\cite{chen,sawatzky,kohno}, no universal operative principle has been enunciated except for the general phenomenon of Mottness\cite{rmpphillips}.  In his parting words in 2016, P. W. Anderson\cite{andersonLW} chided  condensed matter theorists for not facing up to this problem:  ``I remain baffled by the almost	 universal	refusal of	theorists to confront	this	evident fact of hole-particle	
asymmetry head-on.''  It is this task we take on in this paper.  What all cuprates have in common is that the parent material cannot be understood without considering the interactions.  The minimal model thought to be relevant in this context is due to Hubbard in which electrons move on a square lattice but pay an energy cost whenever  opposite-spin electrons reside on the same site.  Since this model is unsolvable in any dimension other than $d=1$, it is difficult to pin-point a clear organizing principle, other than that the interactions are important, as the root cause of the asymmetry.  An added complication is that the Mott insulating state that arises from the local interactions is thought to be featureless above any temperature associated with ordering, just as is the Fermi liquid, the non-interacting limit.  Consequently, appealing to some sort of symmetry breaking appears to be a non-starter.

We propose here that such an organizing principle can be unearthed by focusing on the full symmetry group of a Fermi liquid and analyzing which symmetries in the Fermi liquid survive the transition to the paramagnetic Mott insulator\cite{dmft}. While it is common to use the Hubbard model to study this transition, our key point here is that the essence of the Mott transition is captured by a simpler model which breaks the fundamental local-in-momentum space $\mathbb Z_2$ symmetry ({non-local in real space}) of the Fermi liquid state. This $\mathbb Z_2$ symmetry breaking serves as an organizing principle for Mott physics. We find that both local on-site Hubbard and local-in-momentum (as in the exactly solvable Hatsugai-Kohmoto model\cite{HK,hksupercon,baskaran}) (HK) interactions fall into the same universality class as they both break $\mathbb Z_2$ symmetry.  We then use K-theory to show that the surface of zeros that characterizes the Mott phase is stable to perturbations, thereby establishing the existence of a fixed point.  Our work here is analogous to that of   Ho\v{r}ava's\cite{horava} on the stability of a Fermi surface.  

\section{Relevance of the  HK Interaction}

\begin{figure*}
\centering
\includegraphics[scale=0.7]{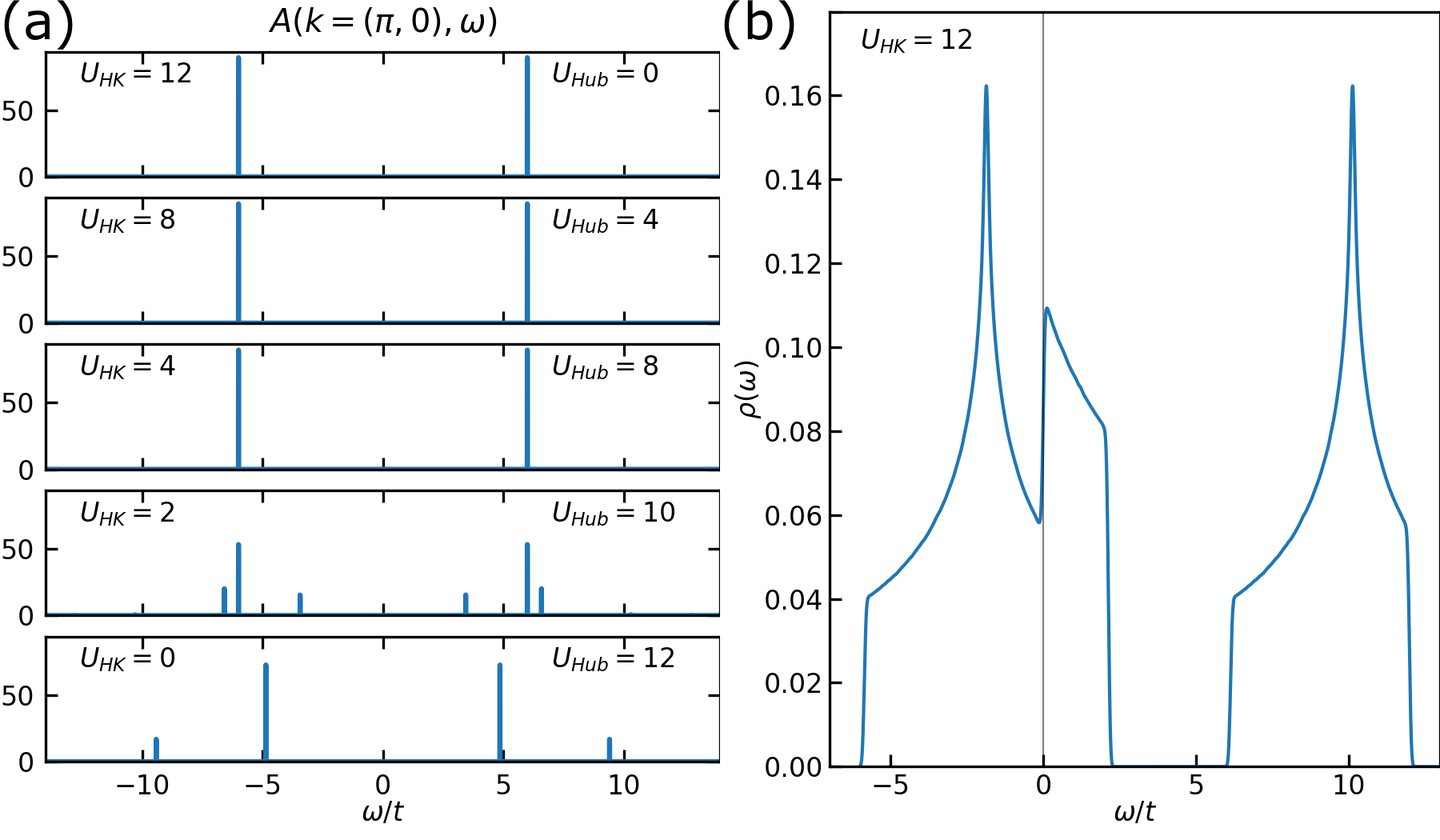}
\caption{
 a.) Spectral function of HK and Hubbard models from exact diagonalization with parameters shown.  At half-filling there is little difference between the models showing that the HK interaction accurately models the U-scale physics of Hubbard.  b.) Density of states of the HK model at filling $\langle n \rangle=0.8$, showing a strong particle-hole asymmetry at the Fermi energy.}
 \label{fig1}
\end{figure*}

Part of the motivation for this work is that there seem to be two disparate ways of generating a Mott transition with no apparent relationship between them.  These constitute the Hatsugai-Kohmoto\cite{HK}(HK) and Hubbard models.  While both models contain the standard kinetic term, the HK model contains a non-standard local in momentum space interaction
\beq
H_{\rm int}^{HK}=U\sum_{k}n_{k\uparrow}n_{k\downarrow},
\eeq
and the Hubbard model, the standard
real-space
\beq
H^{\rm Hubb}_{\rm int}&=&U\sum_i n_{i\uparrow}n_{i\downarrow}\nonumber\\
&=&U\sum_{k,p,q} c^\dagger_{(k-q)\uparrow}c_{k\uparrow}c^\dagger_{(p+q)\downarrow}c_{p\downarrow}
\label{eqhubb}
\eeq
interaction, where we have written the Fourier transform to show the explicit non-local momentum structure.  Even with the kinetic energy, the former model is solvable exactly\cite{HK,hksupercon}, yielding an insulating state with a hard gap should $U>W$, where $W$ is the bandwidth.  The explicit energy cost for doubly occupying the same $k-$state is the explicit mechanism for the Mott physics in the HK model as it leads to singly occupied states below the chemical potential.  Ultimately the same must be true also for the Hubbard model but here only numerics\cite{dmft}  support a gap indicative of Mott physics.  Hence, it is worth comparing both models. Note the $q=0, k=p$ part of the Hubbard interaction is of the same form as the HK interaction. {As will become evident from our analysis, it is this term that is the leading relevant interaction that drives Mott physics.   As shown in Fig. (\ref{fig1}), the spectral functions for both models are roughly identical.  Both describe a gapped state in which spectral weight lies at high and low energies above the gap.  It is the presence of such spectral weight at high and low energies that generates the surface of zeros of the real part of the single-particle Green function, denoted as the Luttinger surface\cite{setty,dzy}. The surface of zeros only appears at momenta which are singly occupied\cite{dzy,HK,hksupercon}. Evident from Fig. (\ref{fig1}a) is that there is no difference between the models in the first  three panels.  Differences only emerge at high enegies but with significantly small spectral weight.}   Nonetheless, that the HK model is capable of capturing Mott physics is not widely appreciated.  Unearthing why these two quite apparently different models yield the same physics is the primary goal of this paper.  An added feature which the spectral functions lay plain is that the real-part of the single-particle Green function evaluated at 

The first thing that must be established with the HK model is why does the interaction $H_{\rm int}^{\rm HK}$ destroy Fermi liquid behaviour.  Two distinct arguments will be adopted here.  First, we appeal to the renormalization principle\cite{polchinski,weinberg} for fermions and show that $H_{\rm int}^{\rm HK}$ is a relevant perturbation.  The correct starting point for renormalization of fermions is to demand that the kinetic term in the action,
\beq	
S_0=\int dtd^d\vec p \psi_\sigma^\dagger(\vec p)(i \partial_t - (\epsilon_{\vec p}-\epsilon_F))\psi_\sigma(\vec p)
\eeq
has zero scaling dimension under the distortion $\vec p=\vec k+s\vec \ell$
where $\vec k$ is along the Fermi surface, $\ell$ is perpendicular to it and $s$ the scaling parameter which will be set to $0$ to preserve the Fermi surface.  Expanding the dispersion relationship of an electron around the Fermi surface,
\beq\label{expansion}
\epsilon(\vec p)=\epsilon_F+\vec \ell\frac{\partial\epsilon}{\partial \vec p}+O(\ell^2),
\eeq
we find that demanding $[S_0]=0$ requires that $\psi_\sigma(\vec p)\rightarrow s^{-1/2}\psi_\sigma(\vec p)$.
 The irrelevance of a generic interaction term
\begin{widetext}
\beq\label{vint}
S_{\rm int}=\int dt \prod_{i=1}^4d^{d-1}\vec k_i d\ell_i V(\vec k_1,\cdots,\vec
k_4)\psi^\dagger_\sigma(\vec p_1)\psi_\sigma(\vec
p_3)\psi^\dagger_{\sigma'}(\vec p_2)\psi_{\sigma'}(\vec
p_4)\delta^d(\vec p_1+\vec p_2-\vec p_3-\vec p_4).\nonumber\\
\eeq
\end{widetext}
follows because $[S_{\rm int}]=1$ (arising from $s^4$ from the four factors of $d\ell$, $s^{-1}$ from $d\t$ and $s^{-4/2}$ from the four fermion fields) and hence vanishes in the $s\rightarrow 0$ limit.  This conclusion holds even if loop corrections with $L$ loops are included as they scale as $s^L$, thereby vanishing for a generic interaction.  Note an interaction of the form $Un_{k\sigma}n_{k'\sigma}$ has a tree-level scaling dimension of $-1$ and hence contributes the same order as does the chemical potential.  That is, it leads to the mass renormalization of Fermi liquid theory.  Loop corrections of this term once again contribute $s^L$ and hence generate no self-energy corrections. Within this scheme, the only exception arises when electrons scatter with momenta on opposite sides of the Fermi surface.  In this case, the $\delta$-function factorizes and the interaction is marginal and leads to an instability should $V<0$.  Note the form of the kinetic energy term is irrelevant to this argument.  All that is necessary is the expansion in Eq. (\ref{expansion}).

Contrastly, the local-in-monentum space interaction
\beq\label{hkint}
\tilde{S}_{\rm int}=U \int dt d\ell d^{d-1}\vec k \psi^\dagger_\uparrow(\vec k)\psi_\uparrow(\vec
k)\psi^\dagger_{\downarrow}(\vec k)\psi_{\downarrow}(\vec k),
\eeq
 differs from the generic interaction in $S_{\rm int}$ in that it contains only a single integration over momentum.  This interaction can be derived from a non-local in space interaction that preserves the  center of mass of interacting pairs of electrons\cite{HK}.   Because of the single integration over momentum in Eq. (\ref{hkint}), the tree-level scaling of $\tilde{S}_{\rm int}$ takes the form $s^{-2}$ and the interaction term is in fact relevant {even if the electrons do not lie on the Fermi surface.} Once again, loop corrections are irrelevant to this term following the argument in Weinberg\cite{weinberg,weinberg2}.   The key conclusion here then is that the interaction in the HK model provides a relevant deformation of Fermi liquid theory.  No contradiction arises from the traditional renormalization principle for Fermions\cite{polchinski,shankar,weinberg,weinberg2} as the HK interaction arises from non-local real-space interactions.

 \section{$\mathbb Z_2$ Symmetry Breaking in Mott Physics}
 
 That the HK interaction provides the general mechanism for the breaking of long-range real-space entanglement of a Fermi liquid, {thereby providing a proxy for the Hubbard model}, we appeal to a little-known observation by Anderson and Haldane\cite{AHO4} regarding the full symmetry group of a Fermi liquid.  
Their key point is that because Fermi liquids possess separately conserved currents for up and down spins, the full symmetry group for each point on the Fermi surface is $O(4)$, the real group of rotations in ~4-space.  The determinant of an $O(4)$ matrix is either $+1$ or $-1$ thus exhibiting the disconnected nature of this Lie group. Namely, the proper group $SO(4)$ where the determinant is $+1$ cannot be continuously deformed into those whose determinant is $-1$.  To understand what remains, we consider the quotient $O(4)/SO(4)$ which is isomorphic to $\mathbb Z_2$.  The $\mathbb Z_2$ arises simply because there are 2 connected components of $O(4)$.  That is, $\pi_0(O(4))\simeq {\mathbb Z_2}$ (here $\pi_p(G)$ is the group of homotopy classes of maps of the p-dimensional sphere to $G$) which is equal to the group consisting of the identity, $I$, and reflections, $R$. A reflection $R$ through a hyperplane is represented by $R_{ij}= \delta _{ij} - n_in_j$ if $(n_0, \cdots , n_3)$ is the orthonormal vector to the hyperplane.  
There is of course a quite distinct ${\mathbb Z_2}$ which lurks because $\pi _1(SO(4))\simeq Z_2$ which tells us that there is a simply connected double cover of $SO(4)$ called $Spin(4)$ (the spin group) which is isomorphic to $SU(2)\times SU(2)$, one $SU(2)$ for the spin and the other for the charge pseudospin, {thereby giving rise to  an equivalence between spin and charge  degrees of freedom in a Fermi liquid.}  As a result, in terms of the particle-hole spinor, $\psi^\dagger_{\vec p}=(c^\dagger_{\vec p\uparrow},c_{-\vec p\downarrow})$, we can write the Hamiltonian for a Fermi liquid as 
\beq
H_{\rm FL}=\sum_{\vec p} \psi_{\vec p}^\dagger (\epsilon_{\vec p}-\epsilon_F)\tau_3\psi_{\vec p}+\cdots
\eeq
which lays plain the inherent $SU(2)$ invariance of the charge sector as proposed initially by Anderson\cite{ASU2} and Nambu\cite{Nambu} and the existence of an infinite number of conserved currents, $n_{\vec p\sigma}$.  Here $\tau_3$ is the $z$-component of the traditional Pauli matrices.  The ellipses stand for any interaction terms that renormalize to zero or terms that contribute at the same level as the chemical potential which lead to the mass renormalization of Fermi liquid\cite{polchinski,shankar,weinberg,weinberg2}.  The   extra $\mathbb Z_2$ symmetry obtains only for the electrons precisely at the Fermi surface. {In fact, while electrons at the Fermi surface have an $SU(2)$ symmetry those away just have a $U(1)$.} As the kinetic energy vanishes for such electrons, extra symmetries emerge.  The relevant symmetry that emerges within $O(4)$ is that the sign of {\it only} one of the spin currents can be changed without any consequence to the underlying theory.  That is, at the Fermi surface,
a particle-hole transformation on one species $c_{p\uparrow}\rightarrow c^\dagger_{p\uparrow}$ or $n_{p\uparrow}\rightarrow 1-n_{p\uparrow}$ but preserving $n_{p\downarrow}\rightarrow n_{p\downarrow}$ can be made with impunity.  The remaining electrons do not enjoy this symmetry. In this sense, the $\mathbb Z_2$ symmetry is emergent in a Fermi liquid as it is exact only at the Fermi surface. In the presence of generic short-range interactions, the precise manifestation of this symmetry is detailed in the Supplementary Materials.  It is this discrete $\mathbb Z_2$ symmetry that a Fermi surface possesses which ultimately accounts for the inherent particle-hole symmetry at low energies.  {Once this symmetry is lost so is the symmetry between particle and addition around the chemcial potential.}

\begin{figure}
\centering
\includegraphics[scale=0.3]{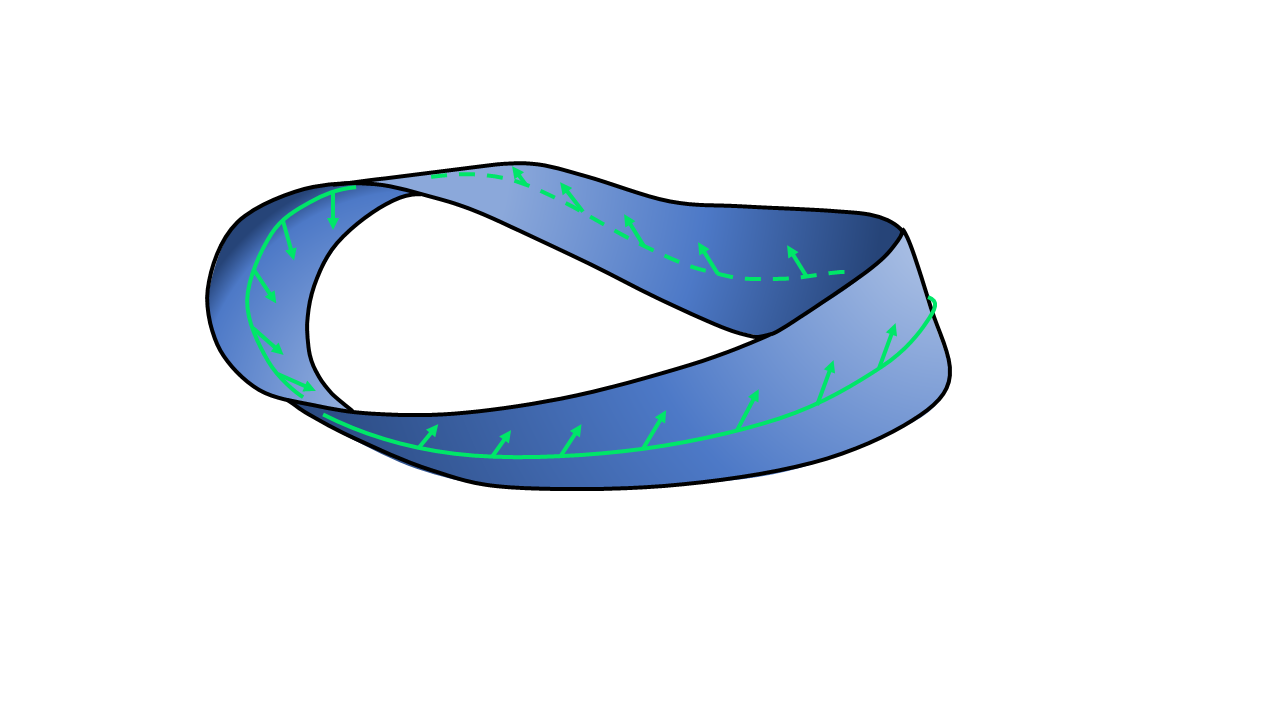}
\caption{
{\bf Mobius bundles:} Parallel transport of the normal vector in the M\"obius bundle along the yellow curve (the Fermi surrface) reveals that going around once leads to a reflection.  A double traversal brings back the original vector, thus laying plain the underlying $\mathbb Z_2$ symmetry.  Note this symmetry is only maintained at the point from which the parallel transport was initiated.
}
\label{mobius}
\end{figure}

There is a subtlety here that points to more than $O(4)$ defining the group structure of Fermi liquids. To establish this, we note that from the $\mathbb Z_2$ symmetry  (of order $\frac{1}{N}$ detailed in the Supplementary Materials), we can view the $O(4)$ action as giving an $O(4)$-bundle structure to the fermions on the Fermi surface.  The $\mathbb Z_2$ symmetry in this context is related to orientability (a consistent orthonormal frame that remains invariant upon parallel transport through a loop as illustrated in Fig. (\ref{mobius})) of the bundle.  The first step\cite{kitaev} is to realize that a Fermi liquid augmented by a number of trivial bands (in a sense we will explain below) has the same properties as the original system. We consider general free Hamiltonians
\beq 
H=\sum_{\sigma, \sigma '}  \psi^\dagger_\sigma (\vec p) A_{\sigma \sigma '} (\vec p)\psi_{\sigma '} (\vec p).
\eeq
We can think of $A_{\sigma \sigma '}$ as a map from the Fermi surface (since we are interested in the $\mathbb Z_2$-symmetry described thus far) to a matrix group and we impose two such $A$'s, say $A_1$ and $A_2$ to be equivalent when
\beq
A_1\sim A_2 \quad \text{ if } A_1\oplus A_{trivial} \sim _{hom}  A_2\oplus A_{trivial} ,
\label{A12}
\eeq
where $A_{trivial}$ represents the trivial system 
\beq
A_{trivial}= \left(
\begin{array}{cc}I  &0\\
0& I\\
\end{array}
\right)|p|^2,
\eeq
and $\sim_{hom}$ means homotopically equivalent. The homotopy equivalence is reflected in being able to continuously deform the eigenvalues without changing the determinant.   
This equivalence class gives rise to the set of maps from the Fermi surface (which we assume to be {\it homotopic} to a sphere) to a classifying space $C_q$ or $R_q$, complex or real, respectively. The only classifying spaces for which $\pi_0(G)\simeq \mathbb Z_2$ corresponds to either $G=O(n)$ or $G=O(2n)/U(n)$ as is evident from the tables in the Supplementary Materials.  This means the additional group $O(4)/U(2)$, {which describes spin-polarized electrons, for example,} is a possible candidate to describe Fermi liquids.  However, such a group would not allow a description in terms of $H_{\rm FL}$.  The types of Fermi liquids described by $O(4)/U(2)$ is beyond the scope of this paper.

\begin{figure}
\centering
\includegraphics[scale=0.35]{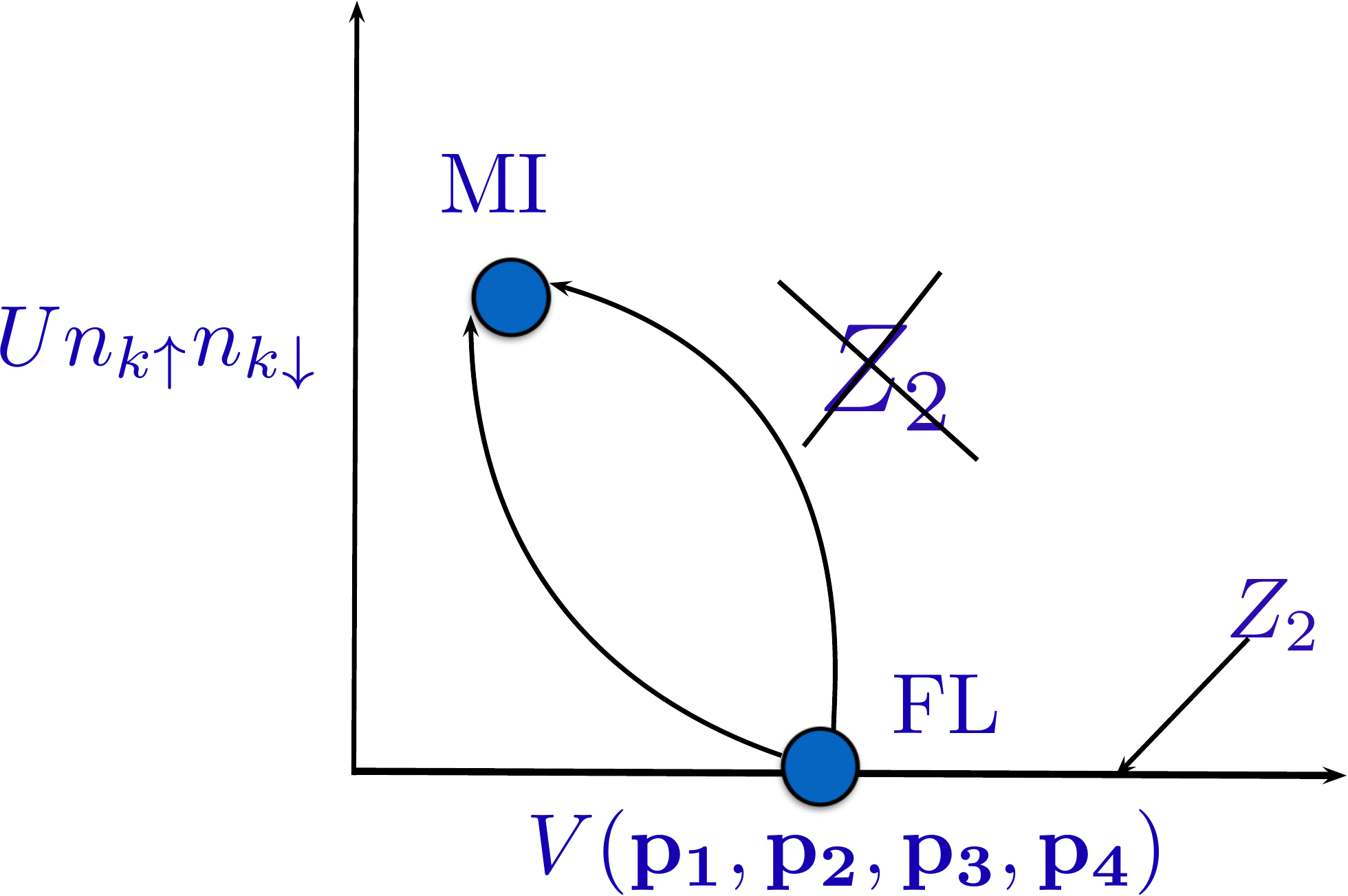}
\caption{
{\bf Hubbard-HK Comparison}   Flow diagram if the interaction in the HK model is made to have zero scaling dimension.  The general 4-momentum interaction has scaling dimension 3 and hence is irrelevant.  FL denotes Fermi liquid which arises strictly along the x-axis which preserves a discrete momentum-space $\mathbb Z_2$ symmetry.  Any non-zero value of $\tilde{S}_{\rm int}$ destroys the Fermi liquid and as a result $\mathbb Z_2$ symmetry.  This scaling analysis is also supported by the exact solution of the HK model.  Note, the Mott insulating behaviour persists even when the generic 4-fermion interaction $V_{\rm int}$ is turned on implying that MI generated from $\tilde{S}_{\rm int}$ is a stable fixed point. 
}
\label{flow}
\end{figure}

From the analysis above, it is clear that any interaction of the form $n_{p\uparrow}n_{p\downarrow}$ (the interaction in $\tilde S_{\rm int}$),  maximally breaks the momentum-space $\mathbb Z_2$ symmetry ($n_{p\uparrow}\rightarrow (1-n_{p\uparrow}), n_{p\downarrow}\rightarrow n_{p\downarrow})$ of a Fermi surface as it transforms to $(1-n_{p\uparrow})n_{p\downarrow}$ and hence the 2-body term changes sign.   Since this term is a relevant perturbation of a Fermi liquid, it is not a surprise that it breaks the $\mathbb Z_2$ invariance of the  would-be Fermi surface. Guided by the $\mathbb Z_2$ symmetry and the principle of relevance, we can analyze the Hubbard interaction as well.  {Explicitly, Eq. (\ref{eqhubb}) tells us that we can organize the Fourier transform as $S_{\rm int}^{\rm Hubb}=S_{3}+S_2+S_1(\tilde{S}_{\rm int}/N)$, where $S_n$ has n-independent momenta. $S_1$ corresponds to the $q=0,p=k$ term, for example which is just $\tilde{S}_{\rm int}/N$ ($N$ the system size) and hence has scaling dimension $-2$.   As each integration over monentum carries with it a power of the scaling parameter $s$,  $S_3$ and $S_2$ are subdominant, ($[S_3]=0$ and $[S_2]=-1$ and hence contributes the same order as the chemical potential) relative to $S_1$, the HK term, which has scaling dimension $-2$.  Because $s$ cannot vainsh faster than $1/N$, the HK term and $S_1$ have identical scaling and hence both are relevant contributions to the Hubbard model.   As seen from the similarity of the gaps in Fig. (\ref{fig1}), we infer that both models are in the same high-temperature universality class.}  This does not mean that extra physics cannot be encoded in the Hubbard model.  As in a Fermi liquid, the Landau interaction parameters can modify the susceptibilities and the density of states and even make the spin and charge sectors differ in 2D\cite{AHO4}.  The key point is that even in the presence of such interactions, the excitations are still governed by the full $O(4)$ symmetry of a Fermi liquid.  Likewise, in both the HK and Hubbard models, the reduced symmetry as a result of breaking of $\mathbb Z_2$ governs the nature of the excitations not the form of the density of states.  That is, the breaking of $\mathbb Z_2$ symmetry by $\tilde S_{\rm int}$ creates a new quartic fixed point as denoted in Fig. (\ref{flow}).  The presence of a charge gap but gapless spin degrees of freedom in the half-filled state are manifestations of the breaking of the discrete $\mathbb Z_2$ symmetry as the spin and charge currents can no longer be rotated freely.  In the doped state, it is well known\cite{sawatzky,rmpphillips,chen,kohno} that the density of states of a doped Mott insulator (see Fig. (\ref{flow})) lacks particle-hole symmetry as must be the case if $\mathbb Z_2$ symmetry in momentum space is absent.  Consequently, both the Mott insulating (gapped charge but gapless charge degrees of freedom) and doped system (absence of particle-hole symmetry) are affected by the breaking of $\mathbb Z_2$ symmetry.  We see then that $\mathbb Z_2$ symmetry is a powerful organizing principle of strongly correlated Mott physics.  This conclusion lends credence to the perturbative result that the gap in Weyl Mott insulators metals with HK interactions is not affected to second order once Hubbard interactions are introduced\cite{wmi}. As a result, we conclude that generically the HK term controls the flow of Fermi liquid to the paramagnetic MI state as depicted in Fig. (\ref{flow}).   Hence, both the HK and Hubbard interactions  break the  $\mathbb Z_2$ symmetry of a Fermi surface and as a result the transition from a Fermi liquid to a Mott insulator involves the breaking of a discrete $\mathbb Z_2$ symmetry.  

  In the spirit of naturalness, it makes sense to scale towards the interactions not the Fermi surface, and hence away from the phase that preserves the $\mathbb Z_2$ symmetry. This causes a major conceptual leap as we can no longer rely on a Fermi surface.  In the presence of strong interactions, it is sensible to choose instead the surface of  zeros of the single-particle Green function, that is, the Luttinger surface, the locus of points in momentum space along which the single-particle Green function vanishes\cite{dzy}. Such a surface demarcates the paramagnetic Mott gap in a single-band system {as numerical simulations lay plain for the 2D Hubbard model\cite{stanescu,imada}}. As shown previously\cite{rosch,dave}, such a surface has nothing to do with the particle density unlike the Fermi surface but rather sets the conditions for spectral weight transfer (SWT) on the Mott scale\cite{dzy}.  Because such SWT\cite{dmft,rmpphillips} is the defining feature of the paramagnetic Mott insulator,  we are interested in a stability anlzysis of this surface in terms of which perturbations destroy it.  As with the Fermi surface, the Luttinger surface can be only be destroyed by perturbations in the perpendicular direction.  For instance, in the HK model, the analogue of the expansion in Eq. (\ref{expansion}) for the Luttinger surface is
\beq
\varepsilon(k)=\frac{U}{2}+\xi(k_L)+\vec \ell\frac{\partial\xi}{\partial \vec p}+O(\ell^2),
\label{luttsuf}
\eeq
where once again, the only degree of freedom is a distortion perpendicular to the surface.  Here $k_L$ is the position of the surface of zeros.  Hence, for the location of the Luttinger surface, $U$ is fixed.  Scaling towards the Luttinger surface is now given by Eq. (\ref{luttsuf}).   If no relevant interactions are found, then this will effectively define a strongly coupled fixed point.  Requiring that $[\tilde{S}_{\rm int}]=0$ fixes the scaling dimension of the Fermion field to be $[\psi_\sigma(p)]=0$ for the HK interaction. Under this scaling scheme,
$[S_{0}]=1$, implying that the kinetic term is irrelevant perturbatively in the Mott insulating state. This reinforces the naturalness of our scheme.  Recall the metallic phase only obtains for $W>U$ and hence cannot be reached perturbatively from the strongly coupled fixed point.  Likewise, the generic four-fermion interaction in Eq. (\ref{vint}) is also irrelevant as it scales as $s^3$ which vanishes when $s\rightarrow 0$.  Hence, even Hubbard physics ( a general 4-momenta term) cannot flow away from the MI point.  Consequently, we argue, as summarized in Fig. (\ref{fig1}a) that $\tilde{S}_{\rm int}$ constitutes a natural fixed point for Mott physics.  Further, we can assess the role of pairing by considering the term
\beq
 H_p = \frac{1}{L^d} \Delta^\dagger \Delta
 \eeq
where $\Delta = \sum_k b_k= \sum_k c_{-k\downarrow} c_{k\uparrow}$.  Since this term has two momenta, the action for this term, $S_p$, has scaling dimension of either $2$ (HK interaction) or $1$  (for the generalized HK interaction) also implying that the Mott insulator is not perturbatively destroyed by pairing, consistent our recent analysis\cite{hksupercon} which shows that only in the metallic phase does superconductivity obtain. 

\section{HK as a Fixed Point}

In this section, we concern ourselves with explaining how the RG flow works for non-local theories and propose a K-theory stability analysis for the underlying fixed point. The notion of renormalizabilty is in general ill-posed as normally stated, as one generally neglects to mention the space of operators within which a theory is renormalizable.   More explicitly, consider a certain theory described by a classically local action $S(\phi_i)$ of some (not necessarily scalar) fields $\phi_1,\cdots ,\phi_n$. One fixes a certain energy scale $\Lambda$ and integrates out fields whose energy is higher than $\Lambda$ so that the effective action $S_\Lambda$ obtains. This is done by integrating out the fields whose frequency $\omega >\Lambda$, thus spitting the fields into high and low frequencies $\phi= \phi_L + \phi_H$ and then integrating 
\beq 
 \int D\phi e^{i S(\phi)}= \int D\phi _L e^{i S_\Lambda (\phi_L)}
 \eeq
 where $S_\Lambda (\phi_L)= -i\log\left( \int  D\phi _H e^{i S(\phi_L , \phi_H)}\right)$. 
 If $S_*$ is a fixed point, one can write 
 \beq 
 S_\Lambda =S_*+ \int d^d x \sum _i g_i \mathcal O _i
 \eeq
 for some {\it local} operators $ \mathcal O _i$ (they are local, despite the integration of high frequency fields, because we focus on fields with $\omega <\Lambda$). The core of renormalization is in the observation that there is a dimension (of operators) $D=D(d, H_0)$ (where $H_0$ is the Hamiltonian of the Free energy), above which the operators are irrelevant, and the number of {\it local} operators $\mathcal O_i$ whose dimension is less than (or equal) to $D$ is finite (this is because classically local operators are polynomials in the fields $\phi$ and their derivative $\frac{\partial ^I\phi}{\partial x^I}$, having used the multi-index notation. Since there are finitely many of these, one can make sense of  such theories.
 The point we want to make is that this makes sense only because we restrict ourselves to a class of operators allowed (in this case the classically local operators $\mathcal O_i$). But this argument can be generalized to a non-local theory theories in real space whose Fourier transform is of course local. Hence, non-locality in real space poses no real hurdle to the renormalization program.  Here locality in momentum space is the standard notion of locality in which position is replaced by momentum.
 
 For the sake of simplicity, we explain this procedure in the case in which the Hamiltonian is $H=H_0 +H_1$ where $H_0$ arises from a classically local operator (such as kinetic energy) and $H_1$ is non-local in real space but its Fourier transform is local as in the example of $\tilde S_{\rm int}$ in the HK model.  We now simply allow operators $\mathcal O_i$ whose Fourier transform is local and can be written as a combination of {\it fundamental} operators; i.e., operators which are either classically local in position space or that are polynomials in the operator components of $H_1$. Since the degree of these polynomials has to be bounded in order for the dimension of the operators to be bounded, there are only finitely many of this latter type as well.
 
  Stability of the Mott fixed point is tantamount to showing that the defining feature of Mott physics\cite{dzy}, the Luttinger surface (defined in Eq. (\ref{luttsuf})) though not necessarily related to the particle density\cite{dave,rosch}, is stable to perturbations, for example the non-HK terms in the Hubbard model. {Recall the surface of zeros is manifest as long as the spectral weight birfurcates as in Fig. (\ref{fig1}).} To this end, we show that the  Luttinger surface, which can be established exactly for the HK model\cite{hksupercon}, under perturbations of the Hamiltonian is determined by Bott periodicity\cite{bott,atiyah} and ultimately K-theory, much the way a Fermi surface is, as shown by Ho\v{r}ava\cite{horava}.  For our purposes, the importance of RG, besides the existence of the fixed point, is that for small values of the parameters $g_i$, the Green function changes continuously by applying perturbation theory to $Z(g_i)= \int D\phi _L e^{i S_\Lambda (\phi_L)}= 
\int D\phi _L e^{i \left( S_*+ \int d^d x \sum _i g_i \mathcal O _i\right)}$.     Consider the Green function
\beq
G(k,\omega)=\langle \psi(0,0)\psi^\dagger(k,\omega)\rangle=\frac{1}{\omega-\xi_k+\Sigma(k,\omega)}
\eeq
for some Hamiltonian which vanishes along a surface of zeros, the Luttinger surface. In a d+1-dimensional $(k,\omega)$ space, we will regard the Luttinger surface, $\Omega$ to have dimension $d-p$ and hence its co-dimension is $p+1$.  Here $\Sigma$ is the exact self energy.   The precise equation denoting the zero surface as in Eq. (\ref{luttsuf}) is determined by the locus of $(k_L,\omega=0 )$ points at which $\Sigma$ diverges.  We assume the fields, $\psi(k,\omega)$ represent complex Fermions consisting of $N$ components.  We consider a point $k_L$ in momentum space which is an element of the Luttinger locus, i.e.  $\Omega:=\{ \det G =0\} $ and consider a ~p-sphere of radius $\epsilon$ centered at a point $k_\perp$ in the normal directions (i.e. $k_\perp$ is in the normal bundle $\nu_\Omega$ to $\Omega$ and we take a fiber of the $\epsilon$-tubular neighborhood of $\Omega$ identified via the exponential map with the $\epsilon$-sphere bundle $S_\Omega(\epsilon)=\{ (k_L, k_\perp)\in \nu_\Omega:\; \vert k_L -k_\perp\vert= \epsilon\} $).   A perturbation which preserves the Luttinger surface moves the zero of $G$ along $k_\perp$.  If not, it  moves it elsewhere in which case the Luttinger surface is destroyed.  We appeal to topology to show that the latter does not obtain.  
At points in $S_\Omega(\epsilon)$ we have that the complex $N\times N$ matrix $G$ is non-degenerate, since by definition the locus of points in momentum space on which it is degenerate is $\Omega$. Therefore, we obtain a continuous analytic map
\beq
\Omega _\epsilon:S_\Omega(\epsilon) \to GL(N,\mathbb C).
\eeq
Here $GL(N,\mathbb C)$ is the group of invertible complex matrices with $N\times N$ entries.  Fixing a point $k_L\in \Omega$, we have that the relevant set is then $S_\Omega(\epsilon)_{k_L}=\{ k_\perp \; \vert k_L -k_\perp\vert= \epsilon\}$ (which, in the language of fiber bundles, is the fiber of $S_\Omega(\epsilon)$ at $\Omega _L$) and this set $S_\Omega(\epsilon)_{k_L}$ is an $S^p$ sphere and the map $\Omega$ at fixed $k_L$ is 
\beq
\Omega' _\epsilon:S^p \to GL(N,\mathbb C).
\eeq
Any deformation $H_M+ g \,H_2$ of the Hamiltonian $H_M$ (here we think of this as the Hamiltonian of the fixed point which exhibits a MI nature) will deform this map $\Omega$ continuously, thus preserving its homotopy class.
Now, the main observation is that if the homotopy class of $\Omega'_\epsilon$ in the $p$-th fundamental group $\pi _p\left(GL(N, \mathbb C)\right)$ is non-zero, the Luttinger surface $\Omega$ must be stable under small deformations. In fact, if the image via $\Omega'_\epsilon$ of $S^p$ were the trivial class, then the map would be homotopic (so continuously deformable ) to a constant map (i.e. mapping the whole of $S^p$ to a constant invertible matrix). But this would mean that  the map $\Omega_\epsilon'$ could be extended to a map from the solid ball $ B_\epsilon (k_L)=\{ (k_L, k_\perp)\in \nu_\Omega:\; \vert k_L -k_\perp\vert= \epsilon\} $ centered at $k_L$ and of radius $\epsilon$ to $GL(N, \mathbb C)$. This is impossible because $\mathcal G$ is degenerate at $k_L$ by definition of the Luttinger surface.  As a result, stability follows.
Higher fundamental groups are notoriously complicated to calculate, but fortunately for classical groups, via the use of Morse theory, R. Bott\cite{bott} was able to prove that they are periodic (and the period depends on the group) 
\beq
\pi _k\left(\lim _\to GL(N, \mathbb C)\right)= \pi _{k+2} \left(\lim _\to GL(N, \mathbb C)\right)
\eeq
and that in particular in the so-called {\it stable} regime or $N$ sufficiently large compared to $p$ ($N>\frac{p}{2}$ suffices), 
\beq
\pi _k\left(\lim _\to GL(N, \mathbb C)\right)=\left\{ \begin{aligned} & 0 &\text{ if } k \text{ is even}\\&\mathbb Z &\text{ if } k \text{ is odd}\end{aligned}\right.
\eeq
We have thus established the fact that 
{\it Luttinger surfaces of codimension p + 1 in momentum-energy space are stable for $p$ odd, and unstable for $p$ even,} much like the Fermi surface case.

Our work here shows that models exhibiting a Luttinger surface, that is, a surface of zeros  ultimately have a rigorous stability condition based in K-theory.  It would be mistaken to associate the winding number $\pi _k\left(\lim _\to GL(N, \mathbb C)\right)$  with the charge density because in the Fermi surface case, the winding number counts the multiplicity of the poles and  because each pole has a quasiparticle interpretation, the winding is equivalent to knowing the charge.  For the zero surface\cite{dave,rosch}, no quasiparticle interpretation of zeros obtains.  Hence, their multiplicity as indicated by the non-trivial winding number $\pi _k\left(\lim _\to GL(N, \mathbb C)\right)$ has no physical significance.  This ultimately sheds light on why deviations from the Luttinger count\cite{luttinger} with the charge density have been so numerous\cite{dave,rosch}.   The existence of our stability condition implies that the details of the underlying Hamiltonian are irrelevant.  The only quantity of relevance is the Luttinger surface.  Consequently, our analysis puts all models with Luttinger surfaces under the same umbrella as they are controlled by a fixed point whose stability is ultimately controlled by K-theory and lack the $\mathbb Z_2$ symmetry of a Fermi surface. The superconducting transition found earlier\cite{hksupercon} should then be a generic feature of this fixed point. It is from the breaking of the discrete $\mathbb Z_2$ symmetry that the particle-hole asymmetry (see Fig. (\ref{flow})) arises naturally, thereby leading to a direct response to Anderson's reproach.  

\textbf{Acknowledgements} 
P.W.P. thanks David Gross for a probing question, asked during a KITP online seminar, regarding renormalization of non-local interactions that ultimately sparked this work, DMR21-11379 for
partial funding of this project and M. Kaplan-Hartnett for assistance with Fig. 2. E.W.H. was supported by the Gordon and Betty Moore Foundation EPiQS Initiative through the grants GBMF 4305 and GBMF 8691.

\textbf{Competing Interests:} The author declares no competing financial or non-financial interests.

\textbf{Data Availability:} The data that support the findings of this study are available from the corresponding author upon reasonable request.

\bibliographystyle{apsrev4-1}

\bibliography{hkrenormalbib}

\clearpage
\onecolumngrid

\section*{Supplementary Materials}
\subsection{$1/N$ expansion}
 
While the $\mathbb Z_2$ symmetry is protected because of the renormalization of short-range interactions to zero at a Fermi surface, this symmetry can also be viewed from the perspective of an underlying $1/N$ expansion.  As pointed out by Shankar\cite{shankar}, the renormalization principle for fermions can be recast as an effective $1/N$ expansion where $N= \frac{K_F}{\Lambda}$ with $\Lambda$ the cut-off.  The effective action is given by 
\beq 
\begin{aligned}
S_F=& \int \frac{d\theta }{2\pi} \int _{[-\Lambda , \Lambda]^{d-1}}\frac{d\vec{k}}{(2\pi)^{d-1}} \int _{-\infty} ^{\infty} \frac{d\omega}{2\pi} \bar \psi (\omega\vec{k} \theta) (i\omega -k) \psi (\omega\vec{k} \theta) \\&-\frac{1}{K_F} \int \bar \psi_j (\omega_4\vec{k} _4\theta) \bar \psi_j (\omega_3\vec{k} _3\theta) F_{ij} \bar \psi_i (\omega_2\vec{k} _2\theta) \bar \psi_i (\omega_2\vec{k} _2\theta) 
\end{aligned}
\eeq
with $F_{ij}$ containing the functional form of the informations.  The $\Lambda \to 0$ (i.e. $N\to \infty$) limit is the Landau Fermi liquid theory.  Therefore, in the large $N$ limit, the $\mathbb Z_2$ symmetry becomes exact; this of course corresponds to the fact that in the large $N$ limit one reduces to the Fermi surface and therefore making precise the heuristics that the symmetry is exact on the Fermi surface.   A consequence of this is that unless the HK interaction is turned on, the RG flow of the Hamiltonian has a fixed point as shown in Fig. 1a.  In the presence of the HK term, the flow is to the new strong-coupling fixed point.

\subsection{K-theory}

We review here the standard classifying scheme\cite{karoubi} in K- theory. 
Because one has that $S^d\sim \mathbb R^d\cup \{\infty\}$, one can calculate such homotopy classes of maps by considering continuous maps from $S^d$ as continuous maps from $\mathbb R^d$ with a boundary condition at $\infty$ (e.g., the maps converge to a fixed matrix, or equivalence class of matrices, as $|x|\to +\infty$). In $K$-theory this corresponds to the fact that 
\beq
KO^{-i}(S^d)= KO^{-i} (\{ pt\}) \oplus  KO^{-i} (\mathbb R^d)= KO^{-i} (\{ pt\}) \oplus  KO^{d-i} (\{ pt\}).
\eeq
The classifying spaces are denoted by $C_q$ in the complex case and $R_q$ in the real case. The index $q$ is an integer taken to be modulo 2 in the complex case and modulo 8 in the real one by Bott periodicity.  Up to homotopy, the classifying spaces are given by

\begin{table}[ht]
\caption{Complex classifying spaces, with $q$ mod. $2$}
\centering
\begin{tabular}{c c c }
  \hline
  q &0& 1 \\
  \hline \hline
 $C_q$ &$\left( U(k+m)/(U(k)\times U(m) )\right) \times \mathbb Z$ & $U(n)$\\
  \hline \hline 
 $\pi _0(C_q)$ & $\mathbb Z$ & 0\\
  \hline
  \end{tabular}
  \end{table}

and the real ones up to $q=4$ (modulo $8$) 

\begin{table}[ht]
\caption{Real classifying spaces up to $4$ mod. $8$}
\centering
\begin{tabular}{c c c c c c}
  \hline
  q &0& 1 &2&3&4\\
  \hline \hline
 $R_q$ &$\left( O(k+m)/(O(k)\times O(m) )\right) \times \mathbb Z$ & $O(n)$& $O(2n)/U(n)$&$U(2n)/Sp(n)$& $\left(Sp(k+m) / Sp(k)\times Sp(m) )\right) \times \mathbb Z$\\
  \hline \hline 
 $\pi _0(R_q)$ & $\mathbb Z$ & $\mathbb Z_2$& $\mathbb Z_2$& 0 &$\mathbb Z$\\
  \hline
  \end{tabular}
  \label{R-Bott-1} 
  \end{table}
 and finally

\begin{table}[ht]
\caption{Real classifying spaces--the rest}
\centering
\begin{tabular}{c c c c }
  \hline
  q &5& 6 &7\\
  \hline \hline
 $R_q$ &$Sp(n)$ & $Sp(n)/U(n)$ & $U(n)/O(n)$ \\
  \hline \hline 
 $\pi _0(C_q)$ & 0&0&0\\
  \hline
  \end{tabular}
  \label{R-Bott-2} 
  \end{table}
  
 In order for our Hamiltonian to homotopically have the $\mathbb Z_2$ symmetry , we need our Hamiltonian to be equivalent in the sense of 
Eq.~(\ref{A12})  to the free Fermi gas  up to $1/N$-corrections.
In other words, because we want to preserve the $\mathbb Z_2$ symmetry (or the full $O(4)$ symmetry),  the classifying space must arise from the real K-theory ones, $R_q$  and further there are only two possibilities for the classifying space. Either $R_q\sim O(n)$ for $n$ large (here $q$ is defined mod. $8$) or $R_q\sim O(2n)/U(n)$.  This gives a complete topological classification of Fermi liquids. As mentioned in the text only $O(4)$ Fermi liquids preserve the structure of $H_{\rm FL}$.
 
To unlock the underlying geometric structure, recall that a  vector bundle $E\to X$ on a topological space $X$ is  said to have {\it real structure} if it is endowed with an anti-linear bundle isomorphism $ i: E \to E$ such that $i^2=id_E$. If the involution is the trivial one then the bundle is simply a bundle over $\mathbb R$. The K-theory of bundle with real structures is denoted by $KO^{i}(X)$ and the one of real vector bundle is denoted by $KR^{i} (X)$. Bott periodicity then says 
\beq
KR^{i+8} (X) \sim KR^i (X) \qquad \text{ and } KO^{i+8} (X) \sim KO^i (X)
\eeq
 The $KO$ groups of a point $X=\{ pt\}$, and therefore of $\mathbb R^n$ since it is contractible, is 
\begin{table}[ht]
\caption{$KO$ and $KR$ Bott periodicity}
\centering
\begin{tabular}{c c c c c c c c c}
  \hline
  i &0& 1&2&3&4&5&6&7 \\
  \hline \hline
 $KO^{-i}$ &$\mathbb Z$ & $\mathbb Z_2$&$\mathbb Z_2$&0&$\mathbb Z$&0&0&0 \\
  \hline
  \end{tabular}
  \label{Bott-K}
\end{table}
In order to calculate the $KR$ groups of a sphere one writes $S^d=\mathbb R^d\cup \{ \infty\}$ and then calculates
\beq
KR^{-i}(S^d)= KR^{-i} (\{ pt\}) \oplus  KR^{-i} (\mathbb R^d)= KO^{-i} (\{ pt\}) \oplus  KO^{d-i} (\{ pt\})
\eeq
and then use the table above. The same equation holds for $KO^{-i}(S^d)$. Note that Bott periodicity for real K-group in table \ref{Bott-K} is the same as the one in tables \ref{R-Bott-1} and \ref{R-Bott-2} .

The geometric meaning of the $\mathbb Z_2$ corresponding to the two sheets of $O(4)$ is related to orientability of the bundle. The structure group of a bundle being $O(n)$ means that he bundle is endowed with a (real) vector bundle metric. If the bundle is orientable, then the structure group (the group where the transition maps are taken) reduces to $SO(n)$ (which corresponds to fixing a quantization). A typical non-orientable bundle is the M\"obius bundle.

\end{document}